\documentstyle[12pt]{article}

\newcommand{\be}{\begin{equation}}
\newcommand{\ee}{\end{equation}}
\newcommand{\bea}{\begin{eqnarray}}
\newcommand{\eea}{\end{eqnarray}}

\newcommand{\av}[1]{\left\langle{#1}\right\rangle}
\renewcommand{\d}{{\rm d}} 

\newlength{\x}\newlength{\xx}
\newlength{\figx}\newlength{\figxx}\newlength{\figy}\newlength{\originx}
\newlength{\originy}\newlength{\placex}\newlength{\placey}

\newcommand{\place}[3]{\setlength{\placex}{#1}\setlength{\placey}{#2}%
\addtolength{\placex}{\originx}\addtolength{\placey}{\originy}%
\makebox[0pt][l]{\hspace*{\placex}\raisebox{\placey}[0pt][0pt]{#3}%
\hspace*{-\placex}}}
\newcommand{\putfig}[5]{\setlength{\x}{\parindent}\setlength{\xx}{\parskip}
\setlength{\figx}{#1}\setlength{\figxx}{\hsize}%
\addtolength{\figxx}{-\figx}\addtolength{\figxx}{-0.1cm}\setlength{\figy}{#2}%
\setlength{\originx}{#3}\setlength{\originy}{#4}\noindent%
\parbox[b]{\figx}{\vspace{\figy}\noindent#5\hfill}}

\begin{document}

\input epsf
\epsfverbosetrue

\title{Bounds on learning in polynomial time}
\author{Heinz Horner and Anthea Bethge\\
Institut f\"ur Theoretische Physik\\
Ruprecht-Karls-Universit\"at Heidelberg}
\maketitle
\vspace{-12pt}
{\noindent \small  Paper presented at the Minerva Workshop on Neural Networks,
Eilat, 25-27 March 1997}
\begin{abstract}
The performance of large neural networks can be judged not only by their storage
capacity but also by the time required for learning. A polynomial learning
algorithm with learning time $\sim N^2$ in a network with $N$ units might be practical
whereas a learning time $\sim {\rm e}^N$ would allow rather small networks only. The
question of absolute storage capacity $\alpha_c$ and capacity for polynomial learning
rules $\alpha_p$ is discussed for several feed-forward architectures, the perceptron,
the binary perceptron, the committee machine and a perceptron with fixed weights in the
first layer and adaptive weights in the second layer. The analysis is based partially
on dynamic mean field theory which is valid for $N\to\infty$. Especially for the
committee machine a value $\alpha_p$ considerably lower than the capacity predicted by
replica theory or simulations is found. This discrepancy is resolved by new
simulations investigating the learning time dependence and revealing subtleties in the
definition of the capacity.
\end{abstract}

\section{Introduction}

Given some neural network architecture and some set of training examples, the question
is not only, whether the network is able to learn the set, but also how many training
cycles are required to do so. Especially for large networks the requirement of
reasonably fast learning algorithms can pose decisive restrictions. To be more
precise, for a network with $N$ nodes a polynomial learning algorithm, with learning
time  growing as $N^2$, might be acceptable for rather large $N$ whereas an exponential
behavior $\sim e^N$ would tolerate small networks only. This question is going to be
analyzed in the following for the task of a random classification (dichotomy) of a
random set of $P$ input patterns in various feed forward networks with $N$ input nodes
and a single threshold output unit. The quantities of interest are the maximal storage
capacity $\alpha_c=P_{max}/N$ and the maximal polynomial capacity $\alpha_p=P_p/N$
with $P_p$ being the maximal size of the training set which can be learned in
polynomial time.

Methods of statistical physics have turned out to be quite useful for the
investigation of various properties of large neural networks. One of the most
prominent examples is the computation of the storage capacity of a simple perceptron
without hidden units by E. Gardner (1988). For unbiased examples $\alpha_c=2$ is
found for $N\to\infty$. Learning can be done by the perceptron
learning rule proposed by Rosenblatt (1962) which determines suitable couplings
for $\alpha<\alpha_c$ after a finite number of learning cycles. Each learning
cycle consists of the presentation of each pattern once and the adjustment
of the couplings for each pattern requires $\sim N$ computations. The total learning
algorithm is therefore polynomial $\sim N^2$ and $\alpha_p=\alpha_c$. An efficient way
to implement such a rule is the adatron described by Anlauf and Biehl (1990).

The question of maximal storage capacity and learning has also been addressed for more
complicated networks, for instance a perceptron with binary weights $W_i=\pm1$
(Kraut and Mezard (1989), Horner (1992), Horner (1993)) and for perceptrons with hidden
layers (Barkai {\em et al.} (1992), Engel {\em et al.} (1992), Priel {\em et al.}
(1994)). For those architectures $\alpha_p<\alpha_c$ is expected. For the
maximal storage capacity, following Gardner's approach (Gardner (1988)), the volume in
the space of synaptic weights $W_i$ compatible with the learning task is computed
within replica theory. The maximal storage capacity is reached if this volume shrinks
to zero. Depending on $\alpha$, a solution with broken replica symmetry can be
found, indicating the decomposition of the available phase space into disjoint
ergodic components. Alternatively within the framework of dynamic mean field theory a
stochastic motion of the weights $W_i(t)$ in the available part of the phase space is
investigated (Horner (1992)). Depending on $\alpha$, diverging time scales can appear.
This indicates again ergodicity breaking and decomposition of the phase space into
disjoint ergodic components. The critical values of $\alpha$ need, however, not be the
same for the two approaches. Both schemes can be generalized to finite temperature
(noise) and for slowly decreasing temperature the above process corresponds to
learning by simulated annealing. For a finite number of temperature steps this
procedure is polynomial $\sim N^2$.

In the following we discuss the question of learning in polynomial time for several
examples. We start with the simple perceptron, briefly describe the ideas behind the
dynamic mean field analysis and turn to the perceptron with binary weights. This is
followed by a discussion of perceptrons with one layer of hidden units. In the
committee machine learning is done for the weights connecting the input nodes with
the hidden units and fixed connections from the hidden units to the output node. The
number of hidden units in this case is assumed to be finite. Another architecture
(coding machine) proposed by Bethge {\em et al.} (1994) has fixed connections in the
first layer which map the input onto a large layer of hidden units. Learning is done
with the weights connecting the hidden units to the output unit.

We focus exclusively on learning a set of unbiased random patterns leaving aside the
most interesting question of generalization ability for patterns
constructed by some other rule.

\section{The Perceptron}

The perceptron can be viewed as the elementary building block of any neural network. 
It has $N$ input nodes and a single output node connected by synaptic weights $W_i$.
The set of patterns $\mu=1\cdots P$ is characterized by its inputs $\xi^\mu_i=\pm1$
and the desired outputs $\zeta_\mu=\pm1$, both chosen randomly with equal probability.
On presentation of pattern $\mu$, the output unit receives a stimulus
\be
h_\mu={1\over\sqrt N}\sum_i W_i\,\xi^\mu_i .
\ee
The learning task is to find weights such that ${\rm sign}(h_\mu)=\zeta_\mu$
for all patterns. Without loss of generality we may choose $\zeta_\mu=1$ for all
patterns and consider the more stringent task $h_\mu>\kappa>0$ together with the
constraint $\sum_i W_i^2=N$. The resulting maximal storage capacity
$\alpha_c(\kappa)$ is (Gardner (1988)) for $N\to\infty$
\be
\frac1{\alpha_c(\kappa)}=\int_{-\infty}^\kappa {\d h \over\sqrt{2\pi}}
\big(\kappa-h\big)^2\,{\rm e}^{-h^2/2}
\ee
and $\alpha_c(0)=2$. At maximal loading, the probability distribution of the stimuli
$h_\mu$ is
\bea
P(h)&=&\frac1P\sum_\mu\delta(h-h_\mu)\\
&=&\frac12\Big\{1+{\rm erf}\frac{\kappa}{\sqrt{2}}\Big\}\delta(h-\kappa)
+\frac1{\sqrt{2\pi}}\,{\rm e}^{-h^2/2}\Theta(h-\kappa).\nonumber
\eea
We are going to use this result later on.

As learning rule (for $\kappa=0$) we have investigated a slightly modified adatron
(Anlauf and Biehl(1990)), where upon presentation of pattern $\mu$ the weights are
modified according to
\be
\Delta_\mu W_i=\frac1{\sqrt{N}}\,\gamma(t)\Big(\kappa(t)-h_\mu\Big)\,
\Theta\Big(\kappa(t)-h_\mu\Big)-\eta\,W_i.
\ee
The learning time $t$ counts the number of learning cycles. The parameters
are allowed to change during the learning process
such that $\gamma(t)\to1$ and $\kappa(t)\to0$ for $t\to\infty$. The last
term ensures the normalization $\av{W_i^2}=1$ of the weights. Simulations yield for
the median $t_{med}$ of the learning time (time required for error free learning
in 50\% of the training sets investigated)
\be
t_{med}\approx\frac{5.5}{2-\alpha}.
\ee
Since each cycle requires $\sim N^2$ computations this algorithm is polynomial.

\section{Dynamic mean field theory}

Dynamic mean field theory was originally introduced by Sompolinsky and Zippelius
(1982) as alternative to the replica theory of spin glasses. It was applied to
learning in perceptrons with binary weights by Horner (1992) and to perceptrons with
hidden units by Bethge (1997).

Applying dynamic mean field theory to learning in neural networks one defines some cost
function depending on the set of patterns and the weights. It corresponds to the
energy in physical systems, and it is chosen to be zero for weights such that all
patterns are classified without errors, and  positive otherwise. The weights $W_i(t)$
are considered as dynamic variables following some stochastic equation of motion, a
Langevin equation for continuous weights or a master equation for discrete weights.
Both equations allow for finite temperatures corresponding to learning with noise.
Learning by simulated annealing is such a process where the temperature is slowly
reduced. One unit of time corresponds to the presentation of $\alpha N$ patterns and
therefore scales as $N^2$. This means that learning $P=\alpha N$ patterns in finite
time yields a polynomial $N^2$ algorithm.

In the limit $N\to\infty$ a mean field approximation becomes exact. The order
parameters of this theory are correlation functions
\be
Q(t_1,t_2)=\frac1N\sum_i\av{W_i(t_1)W_i(t_2)},
\ee
and corresponding response functions, and in addition similar functions for the stimuli
$h_\mu(t)$. The resulting mean field equations are coupled nonlinear
integro-differential equations. In order to follow learning one would have to
solve these equations with some initial condition, for instance randomly chosen weighs.

Assuming, however, the system has reached equilibrium, the
order parameter functions depend on $t=t_1-t_2$ only and the theory simplifies
considerably. The quantity $1-Q(t)$ is a measure of the portion of phase space
explored in time $t$ (note $Q(0)=1$). 

If the system equilibrates in finite time, which is the case at high temperatures,
$Q(t)$ reaches some asymptotic value $Q_a$ and $1-Q_a$ is a measure of the size of the
total accessible part of phase space. Asymptotically  
$Q(t)-Q_a\sim t^{-\nu}{\rm e}^{-t/t_0(T)}$ is found. At some lower temperature
$T=T_f$ a freezing transition is possible with $t_o(T_f)\to \infty$. For $T<T_f$ one
finds
$Q(t)-Q_c\sim t^{-\nu}$ with $Q_c>Q_a$. This means that the accessible part of
phase space can no longer be explored in finite time and the system is no
longer ergodic.

Depending on the behavior of $Q_c(T)-Q_a(T)$ for $T\to T^{-}_f$ one distinguishes
continuous transitions, if $Q_c(T)-Q_a(T)\to 0$, and discontinuous transitions
otherwise. For a system undergoing a discontinuous transition $Q(t)$
shows a plateau near $Q_c(T_f)$ already above the transition $Q(t)$. This means that
the system for time shorter than some $t_c(T)$ explores primarily one of the ergodic
components which strictly forms only below the transition. The whole scenario is
sketched in  Fig.1.

\putfig{14.5cm}{6.5cm}{0cm}{0.5cm}%
{\place{0.95cm}{-10.17cm}{\epsfxsize=347pt\epsfbox{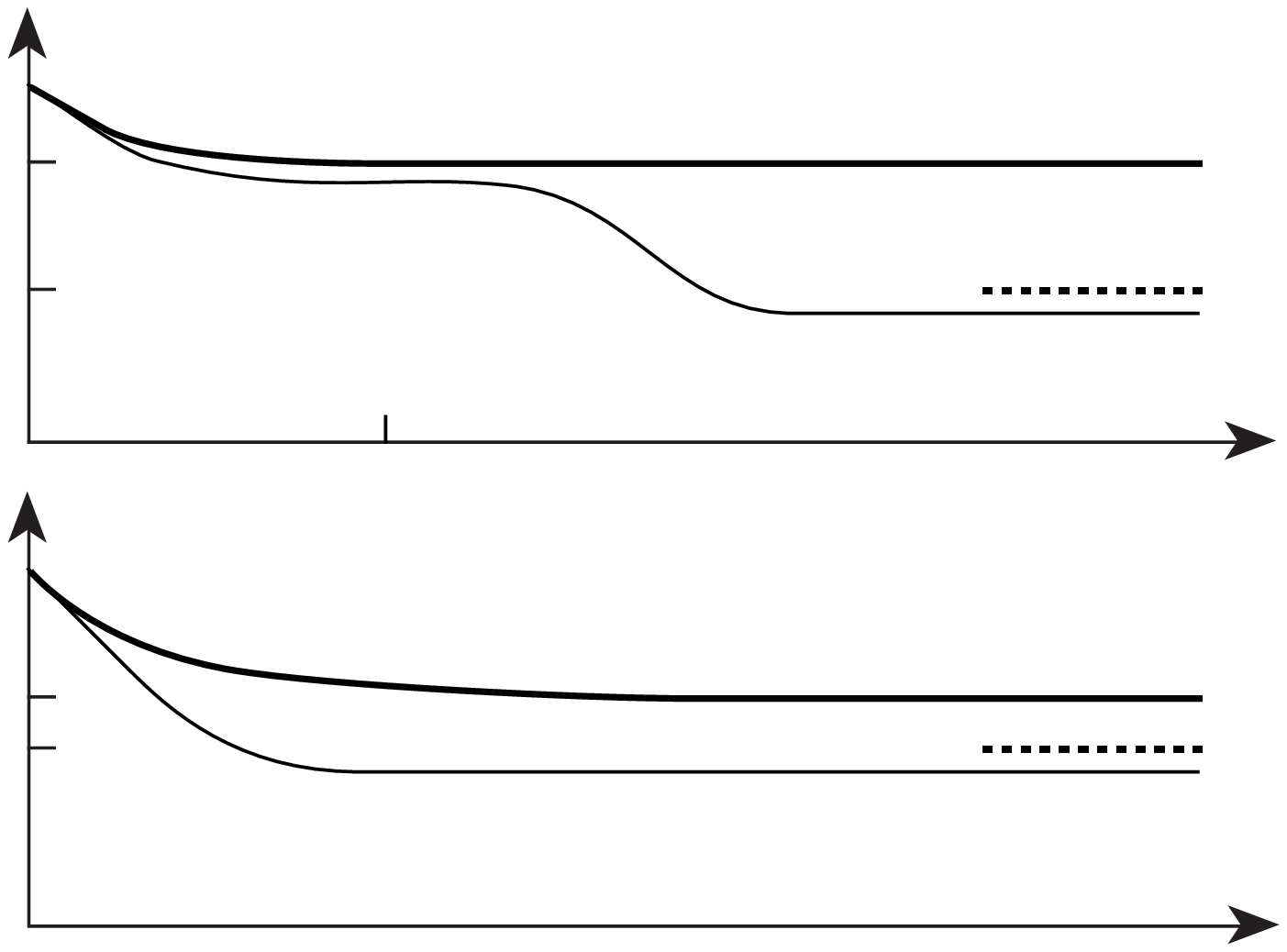}}%
\place{8.95cm}{-10.05cm}{\epsfxsize=347pt\epsfbox{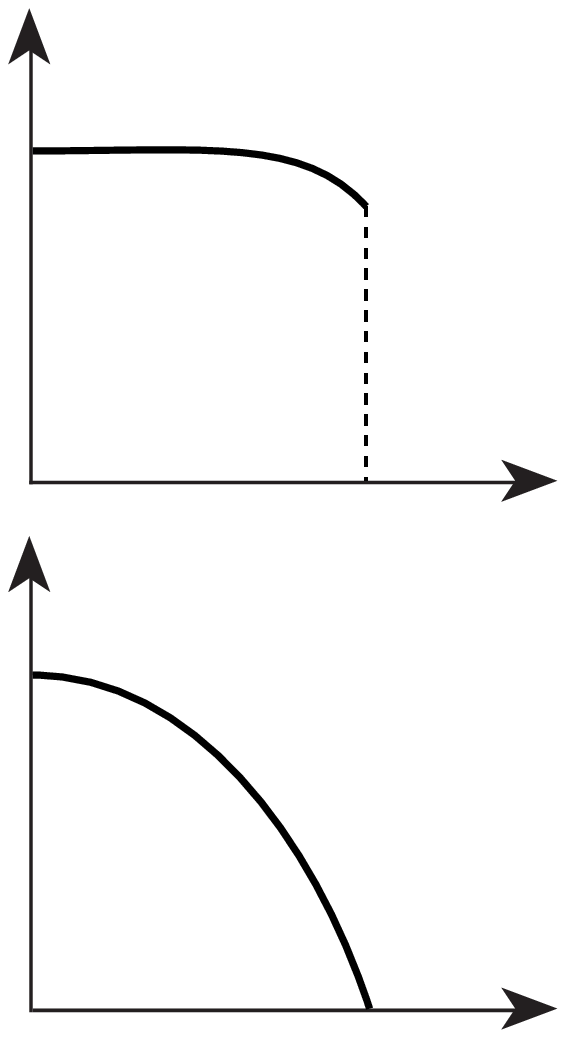}}%
\place{1.3cm}{5.9cm}{$Q(t)$}\place{1.3cm}{2.7cm}{$Q(t)$}%
\place{10.9cm}{5.9cm}{$Q_c-Q_a$}\place{10.9cm}{2.7cm}{$Q_c-Q_a$}%
\place{0.7cm}{5.6cm}{$1$}\place{0.7cm}{2.3cm}{$1$}%
\place{0.6cm}{5.0cm}{$Q_c$}\place{0.6cm}{1.6cm}{$Q_c$}%
\place{0.6cm}{4.2cm}{$Q_a$}\place{0.6cm}{1.1cm}{$Q_a$}%
\place{8.8cm}{3.45cm}{$t$}\place{8.8cm}{0.2cm}{$t$}\place{3.55cm}{3.45cm}{$t_c$}%
\place{13.2cm}{3.45cm}{$T$}\place{13.2cm}{0.2cm}{$T$}%
\place{4.5cm}{5.9cm}{Discontinuous transition}%
\place{5.0cm}{2.4cm}{Continuous transition}%
\place{6cm}{5.35cm}{$T<T_f$}\place{6cm}{1.8cm}{$T<T_f$}%
\place{6cm}{3.75cm}{$T>T_f$}\place{6cm}{0.65cm}{$T>T_f$}}\\
\hspace*{0.5cm}
\parbox[b]{13.1cm}
{\small{\bf Fig.1} Order parameter $Q(t)$ slightly above and below
the freezing transition. Top row: Discontinuous transition. Bottom row: Continuous
transition.\\}

The appearance of diverging time scales for $T<T_f$ is also expected for
nonequilibrium initial conditions appropriate for the question of learning, and one can
therefore conclude that optimal learning is not possible if $T_f>0$. As will be
discussed later this is the case for all values of $\alpha$ for the binary perceptron,
and for $\alpha>\alpha_p$ for the committee machine.

\section{The binary perceptron}

The weights of this network are restricted to $W_i=\pm1$ and learning according to
Eq.(4) is no longer possible. For small networks an exact enumeration of all
values is possible (Krauth and Opper (1989)). Extrapolating to $N\to\infty$ the
value $\alpha_c\approx 0.833$ derived within replica theory (Krauth and Mezard (1989))
is found. This requires, however, $\sim 2^N$ computations.

\putfig{14.5cm}{5.6cm}{0cm}{0.3cm}%
{\place{0cm}{0cm}{\epsfxsize=207pt\epsfbox{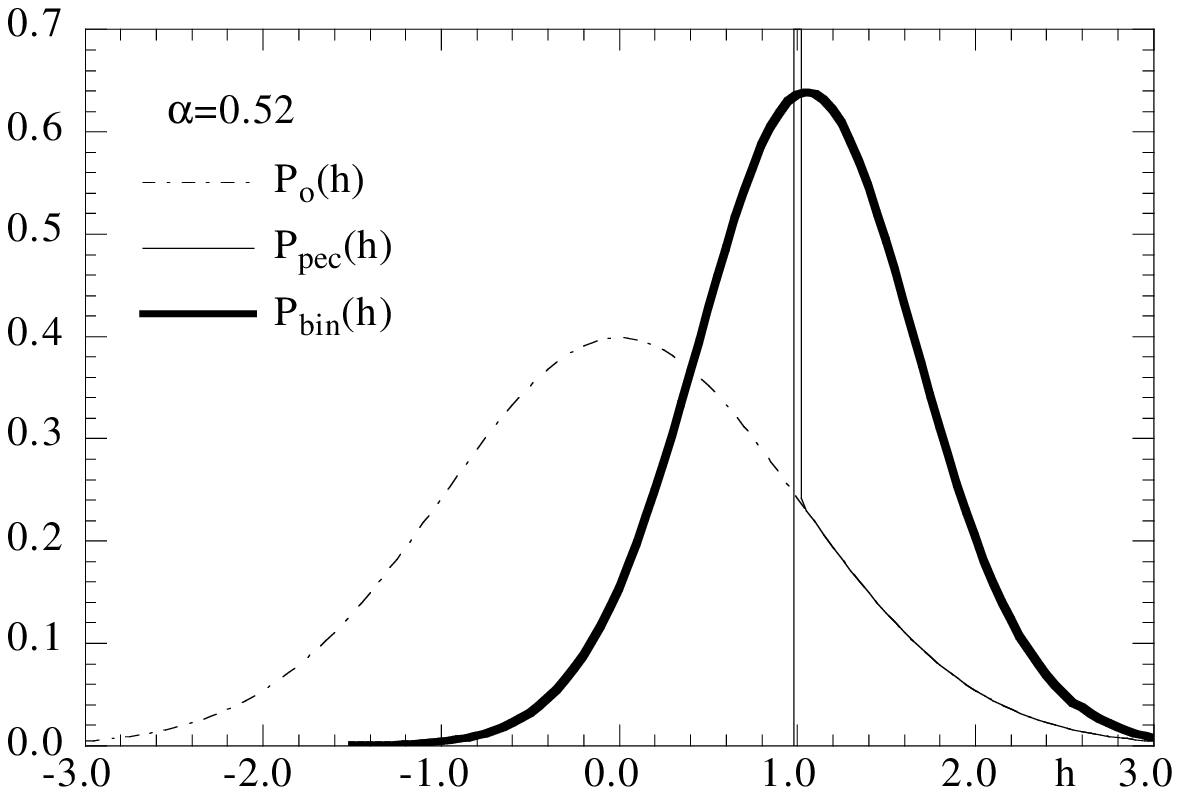}}%
\place{7.45cm}{0cm}{\epsfxsize=207pt\epsfbox{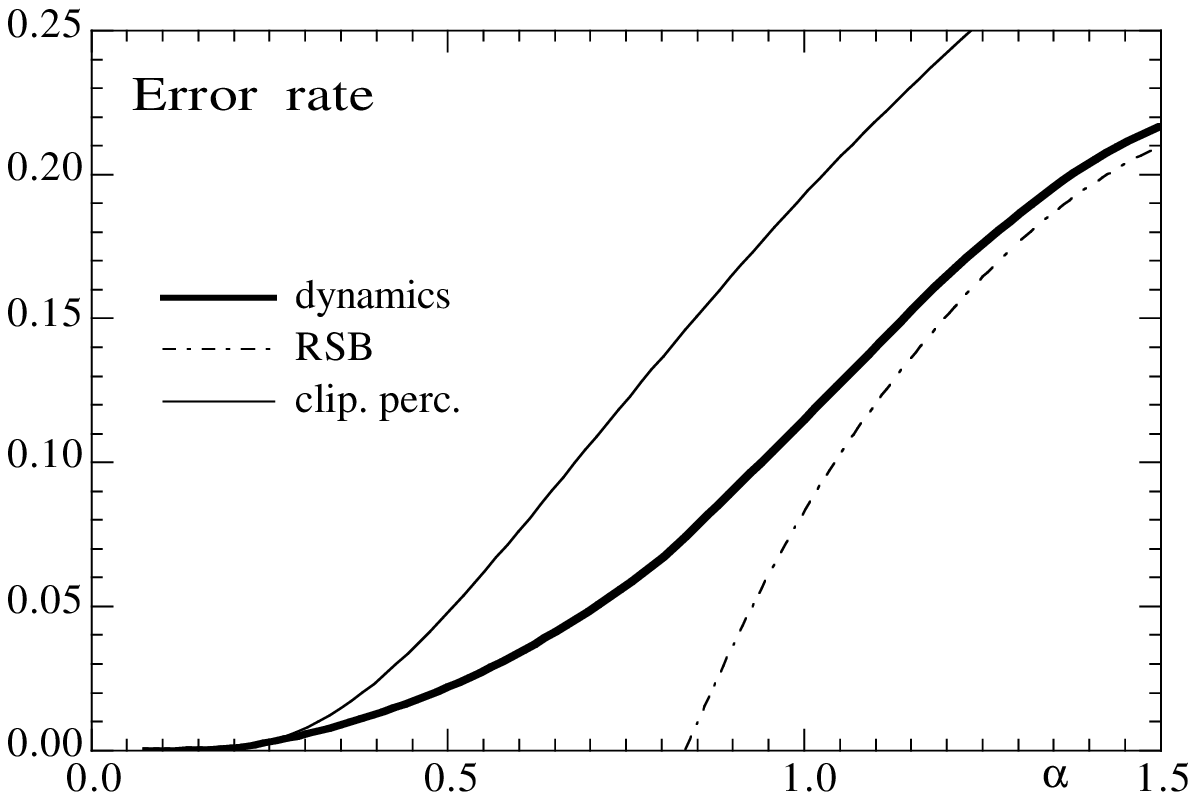}}}\\
\parbox[t]{7.0cm}
{\small{\bf Fig.2} Distribution $P(h)$ for random weights $(P_o)$, perceptron
$(P_{perc})$ and clipped perceptron $(P_{bin})$.\\}%
\hspace{0.7cm}%
\parbox[t]{7.0cm}
{\small{\bf Fig.3} Rate of errors from replica theory (RSB), clipped perceptron
(clip.perc) and dynamics at $T_f$ (dynamic).\\}

Polynomial $\sim N^2$ learning can be done by training according to Eq.(4) assuming
continuous weights $W_i^{(perc)}$ and choosing $W_i={\rm sign}(W_i^{(perc)})$. In
order to estimate the results we can write $W_i=W_i^{(perc)}+\Delta W_i$ and modify
the normalization of the $W_i^{(perc)}$ such that $\av{\Delta W_i}=0$. This yields
$\av{\Delta W_i^2}\approx0.4$ and the distribution of the stimuli $h_\mu$ is obtained
by convolution of the perceptron result, Eq.(3), with a gaussian of width 
$\av{\Delta W_i^2}$. The resulting distribution for $\alpha=0.52$ is shown in Fig.2
and for all values of $\alpha$ there is a tail extending to $h<0$. This means that
part of the patterns are not classified correctly. The resulting error rate is shown
in Fig.3.

Dynamic mean field theory has been applied to this problem by Horner (1992, 1993). A
discontinuous ergodicity breaking transition is found for all $\alpha$ and the
resulting freezing temperature $T_f(\alpha)$ is shown in Fig.4. Simulations with
restricted learning time show that very little improvement of the error rate is
achieved if the system is cooled below the freezing temperature and therefore the
error rate at $T_f$ is a reasonable measure of the performance of an $N^2$
algorithm. As can be seen from Fig.3 the performance is superior to that of the
clipped perceptron, but a finite fraction of errors remains at all values of $\alpha$.

\putfig{7cm}{5.6cm}{0cm}{0.2cm}%
{\place{0cm}{0cm}{\epsfxsize=207pt\epsfbox{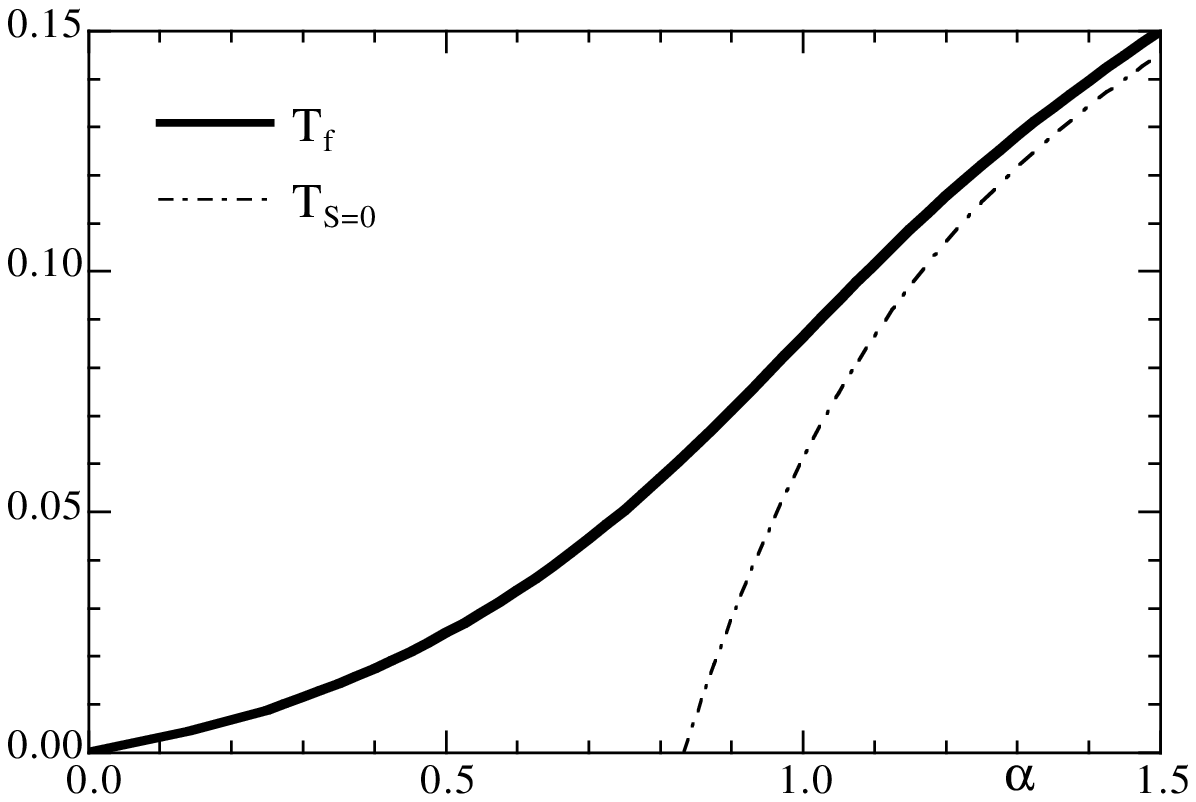}}%
\place{7.4cm}{0.2cm}
{ \ \parbox[b]{7.0cm}
{\small{\bf Fig.4} Dynamic freezing temperature $T_f(\alpha)$ and transition
temperature $T_{S=0}$ derived from replica theory (Krauth and Mezard (1989).}}}

The capacity $\alpha(N,\tau_L)$ for samples of various size $N$ and total learning
time $\tau_L$ has been evaluated by simulated annealing (Horner (1993)). The results
shown in Fig.5 indicate hat the full capacity can be reached for $\tau_L\sim {\rm
e}^N$ only.

\putfig{14.5cm}{6.7cm}{0cm}{0.0cm}%
{\place{0cm}{0cm}{\epsfxsize=255pt\epsfbox{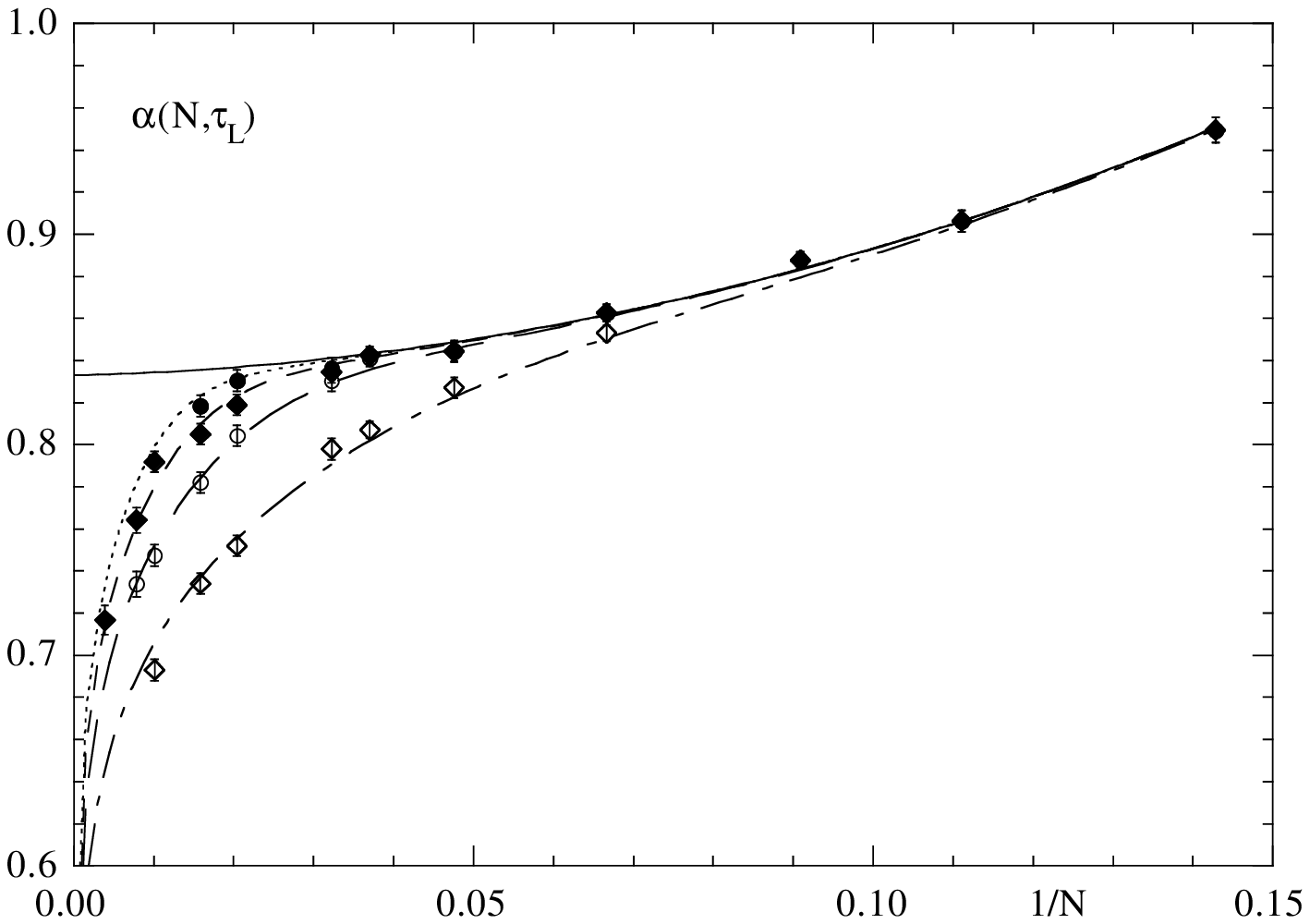}}%
\place{3.5cm}{0.5cm}{\epsfxsize=139pt\epsfbox{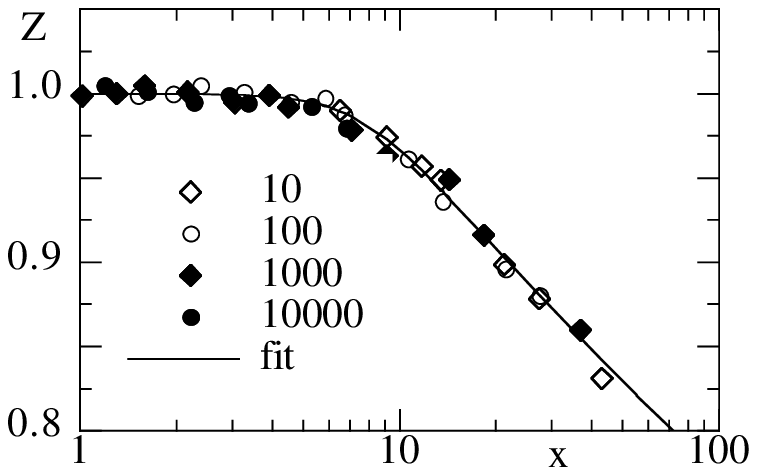}}%
\place{9cm}{0.2cm}
{ \ \parbox[b]{5.5cm}
{\small{\bf Fig.5} Capacity $\alpha(N,\tau_L)$ for the binary perceptron. The insert
shows $Z=\alpha(N,\tau_L)/\alpha(N,\infty)$ with $x=N/{\rm ln}\,\tau_L$.}}}

\section{The committee machine (tree structure)}

The committee machine with nonoverlapping receptive fields (tree structure) has
$N=K\,M$ input nodes, $K$ hidden units and a single output unit. Learning is done
with the weights $W_{i\,l}$ connecting the input nodes with the $K$ hidden units
whereas the weights connecting the hidden units with the output node are fixed $W_l=1$
(see Fig.6). 

\putfig{14.5cm}{4.5cm}{0cm}{0.3cm}%
{\place{-0.1cm}{0cm}{\epsfxsize=202pt\epsfbox{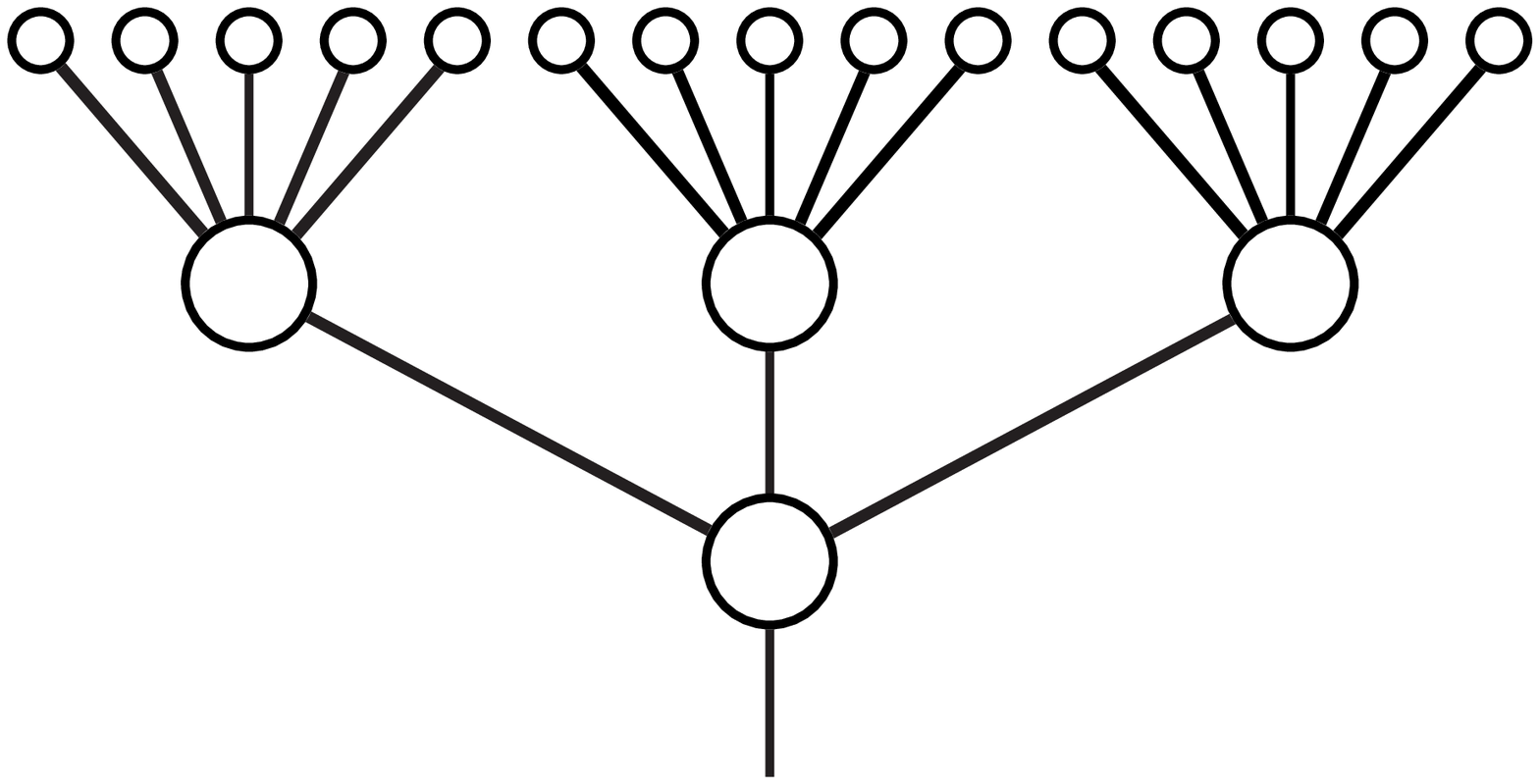}}%
\place{6.7cm}{2.6cm}{$W_{i\,l}$ \ learning}%
\place{6.3cm}{1.2cm}{$W_l=1$ \ \ fixed}%
\place{9.5cm}{0cm}
{ \ \parbox[b]{5cm}
{\small{\bf Fig.6} Committee machine with nonoverlapping receptive fields, $M=5$ and
$K=3$.}}}

\noindent Presenting pattern $\mu$, the hidden units receive a
stimulus, Eq.(1)
\be
h^\mu_l={1\over\sqrt M}\sum_i W_{i\,l}\,\xi^\mu_i .
\ee
With $\zeta_\mu=1$ the learning task is now to determine the weights such that 
\be
\sum_l{\rm sign}(h^\mu_l)>0
\ee

A possible learning procedure is the following: Presenting pattern $\mu$ for example
in a network with $K=3$ one of the possibilities given below shows up during
learning,\\

\hspace*{3cm}
\begin{tabular}{ccccc}
\makebox[0.7cm]{ }&\makebox[0.7cm]{sign$(h_l^\mu)$}&\makebox[0.7cm]{ }& \ prob.  &
learning\\
$+$ & $+$ & $+$ & $1/8$ & $0$ \\
$+$ & $+$ & $-$ & $3/8$ & $0$ \\
$+$ & $-$ & $-$ & $3/8$ & $1 \ {\rm of} \ 2$ \\
$-$ & $-$ & $-$ & $1/8$ & $2 \ {\rm of} \ 3$ \\
\end{tabular}\\

\noindent where equal probability for each possibility is assumed for simplicity.
This means that on average the weights of each subperceptron are updated 
$\frac58\alpha N$ times per learning cycle. Since the perceptron learning rule
allows for each of the 3 subperceptrons a maximal number of $M$ updates at maximal
capacity, we arrive at the estimate $\alpha_p(3)=\frac85=1.6$. This is below the
capacity of a simple perceptron with $N$ inputs and for larger $K$ even lower values
are found, for instance $\alpha_p(5)=1.39$ or $\alpha_p(7)=1.25$.

In general learning for the committee machine can be viewed as a two step
process: i) Selecting the subperceptrons to be modified, ii) modifying the weights of
the selected subperceptrons. The second step is polynomial $\sim N^2$. It might turn
out that the initial restricted random choice of the subperceptrons used to embed a
given pattern is not optimal and that better distributions of the learning load
exist. Testing this possibilities, however, is a combinatorial problem requiring of
the order of $P!$ computations.

A polynomial learning algorithm therefore has to be local in the sense that
the decision which of the subperceptrons to select for training has to be done
instantaneously as learning goes on. In the following we analyze a modified form of
the least action algorithm proposed by Nilsson (1965). Among the candidates for
learning, this means among the subperceptrons with $h_l^\mu<0$, we select those with
the smallest value of $|h_l^\mu|$. This is done by introducing a cost function
$E(\{h_l\})$ which is zero if $\,\sum_l{\rm sign}(h_l)>0$ and monotonously increasing
otherwise. For $K=3$ 
\be
E(\{h_l\})=-\frac1T\, h_1\,\Theta(-h_1)\,\Big(1-\Theta(h_2-h_1)\Theta(h_3-h_1)\Big)\,
+\,{\rm permutations}
\ee
is appropriate. 

With this cost function it is also possible to investigate simulated annealing and to
perform the corresponding analysis based on dynamic mean field theory, which has been
done by Bethge (1997). Her calculation shows that a continuous
ergodicity breaking transition exists for $\alpha>\alpha_p(K)$ with
$\alpha_p(3)\approx 1.75$. The same value was found by Barkai {\em et al.} (1992) and
by Engel {\em et al.} (1992) for the onset of replica symmetry breaking. This
agreement is actually expected for a continuous transition in contrast to a 
discontinuous transition as pointed out in the previous section. Applying the
arguments given in Sect.3, we have to conclude that error free polynomial learning is
not possible for
$\alpha>\alpha_p(K)$.

This appears to be in clear contradiction to the results of simulations reported by 
Barkai {\em et al.} (1992), Engel {\em et al.} (1992) and Priel {\em et al.} (1994)
where for $K=3$ values of $\alpha_c$ between 2 and 2.75 were obtained. Up to 6000
learning cycles were used, a value much larger than what would be expected from
Eq.(5) assuming that the learning time is ruled by the perceptron learning part of
the algorithm. The apparent improvement can also not be due to a better handling of
the combinatorial part where the subperceptrons to be trained are selected. This would
result in a strong size dependence which was not observed. In these investigations
successful learning of all patterns with probability $1/2$ was used as criterion for
$\alpha_c$. Priel {\em et al.} (1994) also determined the median of the time required
for perfect learning as function of $\alpha$. They found for instance for $K=3$ and
$\alpha=2$ the value $\tau_{med}\approx 25$ and a divergence around $\alpha_c\approx
2.75$.

In order to understand this discrepancy we have performed preliminary simulations on
a committee machine with $N=150$ and $K=3$ using the modified least action algorithm
described above together with the perceptron learning rule Eq.(4). Fig.7 shows the
probability $R(\alpha,t_L)$ that after time $t_L$, this means after presentation
of $t_L\alpha N$ patterns, error free learning is not yet reached. For comparison the
learning curve for a simple perceptron is also shown. The results were obtained for a
single set of patterns and in each of the 1000 runs used for each value of
$\alpha$ the patterns were presented in a different random order. For different sets
of patterns similar curves are obtained but we have not tried to average over
different sets or to investigate finite size effects. Judging from the simulations
performed by Priel {\em et al.} (1994) $N=150$ seems to be sufficient to eliminate
drastic finite size effects.

\putfig{14.5cm}{7cm}{0cm}{0.3cm}%
{\place{0cm}{0cm}{\epsfxsize=263pt\epsfbox{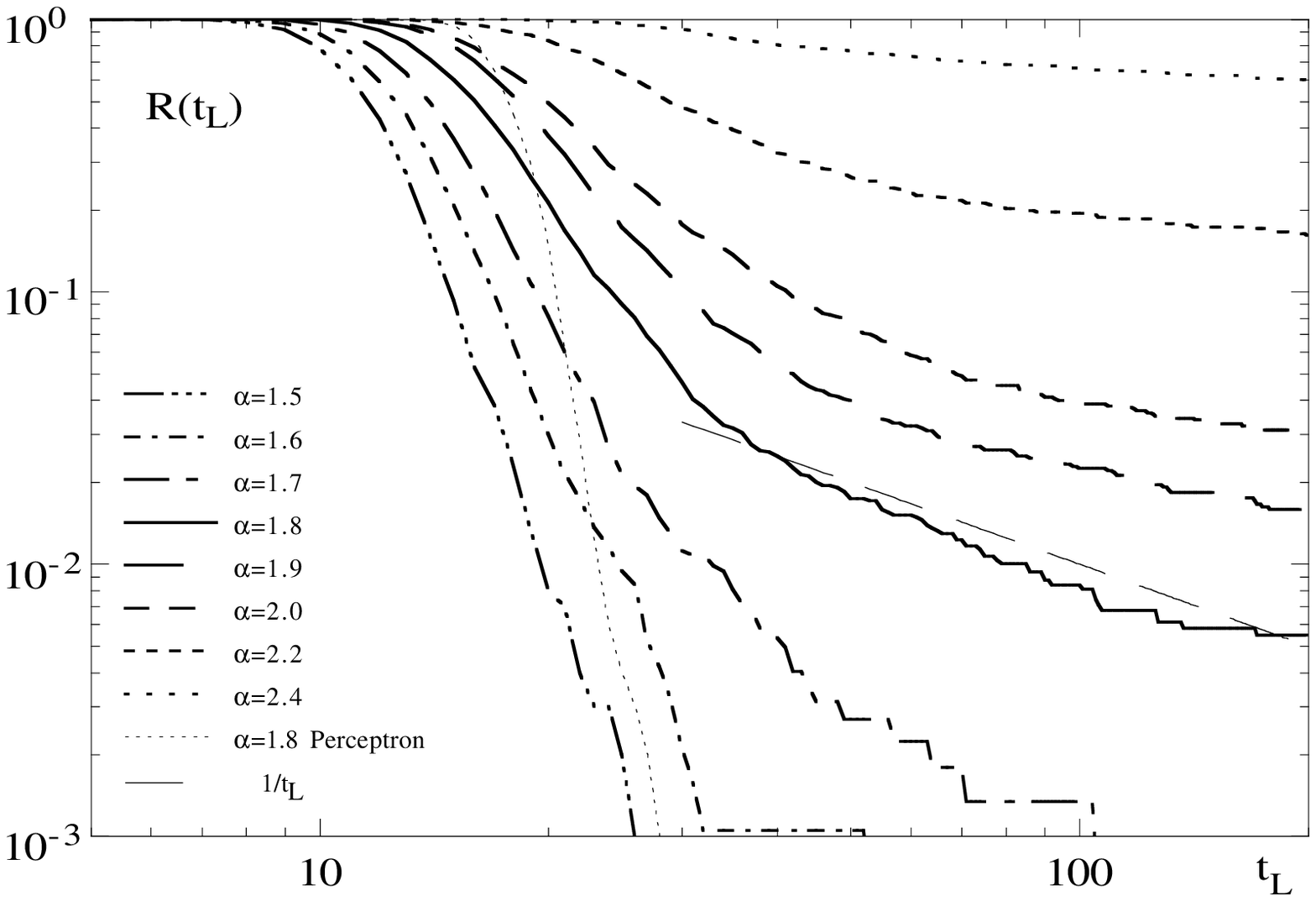}}%
\place{9.3cm}{0.4cm}
{ \ \parbox[b]{5.2cm}
{\small{\bf Fig.7} Fraction of incomplete learning $R(t_L)$ as function of the
learning time $t_L$ for a committee machine with $N=150$, $K=3$ and various values
of $\alpha$.}}}

There appears to be a qualitative difference between the learning curves for
$\alpha<\alpha_p=1.75$ and $\alpha>\alpha_p$, respectively. The fraction
$R(\alpha,t_L)$ decays rapidly to zero for $\alpha<\alpha_p$ whereas a much slower
decay $\sim t_L^{-1}$ or even slower is observed otherwise. For all values of $\alpha$
investigated, a finite median $t_{med}$ of the learning time defined by
$R(\alpha,t_{med})=\frac12$ exists. The average learning time is computed from the
probability $P(\alpha,t_L)=-\partial R(\alpha,t_L)/\partial t_L)$ resulting in
\be
\av{t_L}=\int_0^\infty \d t\,R(\alpha,t).
\ee 
The simulations indicate that for $\alpha>\alpha_p$ this integral does not exist and
furthermore for any finite $t_L$ the probability for error free polynomial learning is
less than one.

This result has certainly to be reconfirmed by additional simulations. Nevertheless it
reveals new subtleties in the determination or even the definition of the storage
capacity of neural networks. If for instance a certain fraction of successful
learning is used to determine the storage capacity different values may result even
in the limit $N\to\infty$  depending on which fraction is required. There could also be
another critical value $\alpha_c$ separating a region where $R(\alpha,t_L)\to0$ for
$t_L\to\infty$ from a region where this limit is finite. The results of the simulation
also allow for speculations on the fractal structure of the accessible part of phase
space being composed of almost disconnected subregions.

\section{Perceptron with divergent preprocessing}

The last architecture to be discussed shows a possibility to increase the storage
capacity beyond the capacity of a simple perceptron yet retaining a polynomial learning
rule. The architecture of this network (coding machine) is shown in Fig.8. The input
layer is divided into $L$ nonoverlapping receptive fields of size $M=N/L$. Each
receptive field has its own part of size $K>M$  of the hidden layer. The fixed weights
connecting input and hidden layer are supposed to establish a one to one mapping of
the input patterns
$\xi^\mu_i=\pm1$ onto internal representations $\hat\xi^\mu_k=\pm1$ assuming
binary inputs and threshold hidden units. For example each of the $2^M$ different
inputs at any of the receptive fields can be mapped for $K=2^M$ onto an internal
representation with a single active node in each part of the hidden layer, generating
a sparse coding internal representation. This mapping could be achieved by unsupervised
learning of the winner takes all type. For $K\ll K_{max}=2^M$ randomly chosen weights
$W_{i\,l}$ would also be possible. If the set of patterns has some structure reflected
onto the subdivision into receptive fields, other unsupervised learning procedures
are possible. The complete hidden layer is finally connected to an output unit and
the corresponding weights $W_l$ are determined by supervised perceptron learning.

The whole architecture can also be viewed as a prototype for data processing in the
brain, having in mind for instance the mapping of a comparatively small number of
neurons in the retina onto a much larger number of neurons in the primary visual
cortex.

\putfig{14.3cm}{6.3cm}{1cm}{1.3cm}%
{\place{-1.9cm}{0cm}{\epsfxsize=309pt\epsfbox{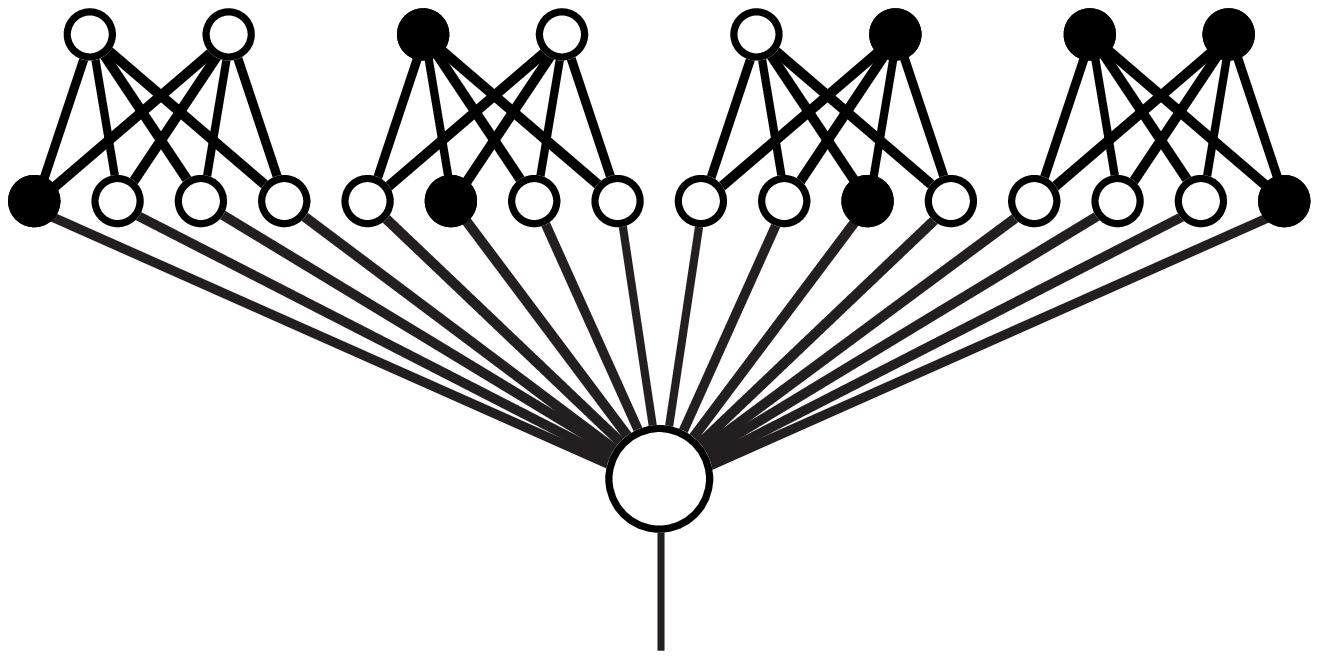}}%
\place{9.95cm}{3.9cm}{$W_{i\,l}$ \ \ fixed}%
\place{9.5cm}{3.5cm}{\scriptsize (unsupervised learning)}%
\place{8.2cm}{1.9cm}{$W_l$ \ \ supervised learning}%
\place{0cm}{-1cm}
{ \ \parbox[b]{12cm}
{\small{\bf Fig.8} Coding machine with $L=4$ receptive fields of size $M=2$ and
$K=2^M$ hidden units for each receptive field.}}}

Even for uncorrelated random input patterns the resulting internal representations are
correlated. Neglecting these correlations for a moment the second layer can be viewed
as a simple perceptron of size $\bar K=L\,K$ and the capacity is $\alpha_p=\alpha_c=2
\, K/M$. Since learning in the second layer can be done with the perceptron algorithm
it is polynomial $\sim (\alpha_c M)^2$. The effect of correlations within the internal
representations has been examined for $K=2^N$ by Bethge {\em et al.} (1994) with the
result
\be
\alpha_c=\frac{2}{M}\Big(2^M-1\Big).
\ee
This shows that the correlations indeed give rise to negligible corrections and the
coding machine  is a useful and fast learning architecture if capacities 
beyond the limits of the perceptron are required.

\section{Outlook}

The performance of different feed forward neural network architectures under the
constraint of learning in polynomial time has been subject of this study. The main
results are the following: For the simple perceptron and the coding machine learning
uncorrelated random patterns the full capacity $\alpha_c$ can be exhausted by a
polynomial learning rule. For a perceptron with binary weights error free polynomial
learning is not possible. The finite size scaling inferred from simulations indicates
that the full capacity $\alpha_c\approx0.833$ obtained from replica theory can only be
reached allowing for learning times $\sim {\rm e}^N$. A mean field analysis of this
architecture yields a discontinuous ergodicity breaking transition in a region where
the replica symmetric solution is stable. 

For the committee machine with nonoverlapping receptive fields a continuous transition
shows up for $\alpha>\alpha_p$, and because of the diverging time scales at the
transition $\alpha_p$ is a bound of the maximal capacity which can be reached  by
polynomial learning. This bound is well below the corresponding value of a simple
perceptron of same size and the value $\alpha_c$ obtained from replica theory with one
step replica symmetry breaking is even higher. The dynamic transition coincides,
however, with the onset of replica symmetry breaking. From simulations, storage
capacities well above $\alpha_p$ were deduced. Preliminary simulations evaluating the
probability of perfect learning as function of learning time indicate that this
discrepancy is due to subtleties in the evaluation of the numerical data. 

We have not addressed the most interesting question of generalization ability for
patterns created by some rule, for instance a teacher neural network. If teacher and
student have the same architecture the situation is quite different, but for different
architectures or for a restricted training set the present discussion might be of
relevance.

\newpage

\section*{References}
\begin{list}{}{\itemsep -2pt\itemindent-28pt}

\item J. K. Anlauf and M. Biehl (1990), Europhys.Lett. {\bf 10}, 387.
\item E. Barkai, D. Hansel and H. Sompolinsky (1992), Phys.Rev.A {\bf45}, 4146.
\item A. Bethge, R. K\"uhn and H. Horner (1994), J.Phys.A {\bf 27}, 1929.
\item A. Bethge (1997), Thesis, Heidelberg.
\item A. Engel, H. M. K\"ohler, F. Tschepke, H. Vollmayr and A.Zippelius (1992),
Phys.Rev.A {45}, 7590.
\item E. Gardner (1988), Europhys.Lett. {\bf 4}, 481.
\item H. Horner (1992), Z.Physik B {\bf86}, 291.
\item H. Horner (1993), Physika A {\bf200}, 552.
\item W. Krauth and M. Mezard (1989), J.Phys.(Paris) {\bf50}, 3054.
\item W. Krauth and Opper (1989), J.Phys.A {\bf 22}, L519.
\item N. J. Nielsson (1965), {\em Learning Machines} (McGraw-Hill, New York).
\item A. Priel, M. Blatt, T. Grossman, E. Domany and I. Kanter (1994), Phys.Rev.E {\bf
50}, 577.
\item F. Rosenblatt (1962),{\em Principles of Neurodynamics} (Spartan, New York)
\item H. Sompolinsky and A. Zippelius (1982), Phys.Rev.B {\bf 25}, 6860.
\end{list}

\end{document}